\documentclass[fleqn,twoside]{article}
\usepackage{espcrc2}
\usepackage{graphicx}

\newcommand{\AmS}{{\protect\the\textfont2
  A\kern-.1667em\lower.5ex\hbox{M}\kern-.125emS}}

\title{Study of the hadronic production of kaon pairs below the threshold for the $\phi$ meson}
\author{P. Moskal\address[cracow]{{Institute of Physics, Jagiellonian University, PL-30-059 Cracow, Poland}}%
\address[julich1]{IKP \& ZEL, Forschungszentrum J\"ulich, D-52425 J\"ulich, Germany},
 M.~Silarski\addressmark[cracow],
 A.~Budzanowski\address[bronowic]{Institute of Nuclear Physics, PL-31-342 Cracow, Poland},
 E.~Czerwi\'nski\addressmark[cracow]\addressmark[julich1],
 R.~Czy\.zykiewicz\addressmark[cracow],
 D.~Gil\addressmark[cracow],
 D.~Grzonka\addressmark[julich1],
 M.~Janusz\addressmark[cracow]\addressmark[julich1],
 L.~Jarczyk\addressmark[cracow],
 B.~Kamys\addressmark[cracow],
 A.~Khoukaz\address[munster]{IKP, Westf\"alische Wilhelms-Universit\"at, D-48149 M\"unster, Germany},
 P.~Klaja\addressmark[cracow]\addressmark[julich1],
 W.~Oelert\addressmark[julich1],
 C.~Piskor-Ignatowicz\addressmark[cracow],
 J.~Przerwa\addressmark[cracow]\addressmark[julich1],
 B.~Rejdych\addressmark[cracow],
 J.~Ritman\addressmark[julich1],
 T.~Sefzick\addressmark[julich1],
 M.~Siemaszko\address[katowice]{Institute of Physics, University of Silesia, PL-40-007 Katowice, Poland},
 J.~Smyrski\addressmark[cracow],
 A.~T\"aschner\addressmark[munster],
 P.~Winter\addressmark[julich1],
 M.~Wolke\addressmark[julich1],
 P.~W\"ustner\addressmark[julich1],
 M.~J.~Zieli\'nski\addressmark[cracow],
 W.~Zipper\addressmark[katowice],
 J.~Zdebik\addressmark[cracow]
}
       
\begin{document}

\begin{abstract}
    The near threshold production of $K^{+}K^{-}$ pairs in proton-proton 
    collisions has been investigated
    at the cooler synchrotron COSY below and above the threshold for the 
    $\phi$ meson using the COSY-11 and ANKE  facilities, respectively.
    The excitation function determined for the $pp\rightarrow ppK^{+}K^{-}$ 
    reaction revealed a statistically significant enhancement close to the threshold
    which may plausibly be assigned to the influence
    of the $K^{-}p$ interaction. In addition, observed consistently by both groups,
    a strong enhancement at low values of the ratio of the $K^{-}p$ to  $K^{+}p$ invariant mass distributions 
    shows that the proton interacts much stronger with $K^{-}$ than with $K^{+}$.
    In this report we focus on  the measurements performed by the COSY-11 collaboration.
    We explain the experimental method used and present  main results of 
    completed analyses
    as well as a new 
    qualitative elaboration of the $ppK^{+}K^{-}$ events population on the 
    Goldhaber plot. We conclude with the observation that 
    event densities 
    increase at the region where the influence from the $K^{+}K^{-}$ 
    interaction is expected. 
  \vspace{1pc}
\end{abstract}

\maketitle

\section{Introduction}
A primary motivation for measuring cross sections for the $pp\to ppK^{+}K^{-}$ reaction near the kinematical threshold
was the study of the hadronic interaction between $K^{+}$ and $K^{-}$ mesons in order to understand the structure of the
scalar resonances $f_{0}(980)$ and $a_{0}(980)$~\cite{walter}. 
Such measurements have been 
made possible by beams of low emittance and small
momentum spread available at storage ring facilities and in
particular at the cooler synchrotron COSY placed in J{\"u}lich, Germany~\cite{cosy}.
A precise determination of the collision energy, in the order of  fractions of MeV,
permitted to deal with  the rapid growth of cross sections~\cite{review} and thus to
take advantage of threshold kinematics like e.g.\ full space phase coverage 
achievable with dipole magnetic spectrometers
rather limited in geometrical acceptance.
Early experiments on $K^{+}K^{-}$ pair production at COSY conducted by the COSY-11 collaboration revealed, however,  that 
the total cross section at threshold is by more than seven orders of magnitude smaller than the total
proton-proton production cross section making the study difficult 
due to low statistics~\cite{magnus,cosy1,cosy2}. 
A possible influence from the $f_{0}$ or $a_{0}$ mesons 
on the $K^{+}K^{-}$ pair production
appeared to be too weak to be distinguished 
from the direct production of these mesons based on the COSY-11 data~\cite{cosy1}.
Recent results obtained by the ANKE collaboration with much higher statistics
can also be explained without the need of referring to the scalars 
$f_{0}$ or $a_{0}$~\cite{anke,c_wilkin}.
However, the systematic collection of  data below~\cite{magnus,cosy1,cosy2} 
and above~\cite{anke,disto} the $\phi$ meson threshold
combined together reveal a significant signal in the shape of the excitation function
in which  the $K^{-}p$ and perhaps also the $K^{+}K^{-}$ interaction manifests itself.
These observations motivate us  
to search for a signal from the interaction 
between the $K^{+}$ and $K^{-}$ in  two dimensional invariant mass distributions.
The analysis is based on generalizations of the Dalitz plot for four particles proposed by 
Goldhaber et al.~\cite{goldhaber1,goldhaber2}.
The knowledge about the KK and KN interactions is important in many physical fields.
In addition to the already mentioned studies of the 
nature of the scalar resonaces $a_{0}(980)$ and $f_{0}(980)$, 
in particular for their interpretation as  $K\bar{K}$ molecules~\cite{hanhart,1,2},
it is also of importance 
in view of discussions on the structure of 
the excited hyperon $\Lambda(1405)$, since it is not clear whether it is a usual  
three quark system or whether it is a  $\bar{K}N$ bound state~\cite{3}. 
Furthermore, an understanding of kaon and antikaon interactions 
with a nucleon is essential for studies of properties of strange particles
immersed in dense baryonic matter~\cite{moskal1} and 
in the determination of the structure of neutron stars~\cite{4,5}.

\section{Measurements of the $pp\rightarrow ppK^{+}K^{-}$ reaction at COSY-11}
\begin{figure}[h]
  \includegraphics[angle=270,width=0.45\textwidth]{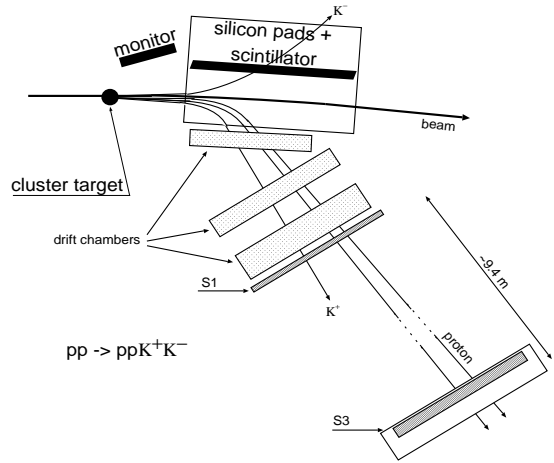}
  \caption{Schematic view of the COSY-11 detector with an exemplary 
    event of the reaction channel $pp\rightarrow ppK^{+}K^{-}$.
    For the description see text.
    \label{detector}
  }
\end{figure}

The measurements of the $pp\rightarrow ppK^{+}K^{-}$  reaction close to threshold
have been conducted using the cooler synchrotron COSY~\cite{cosy} and the 
COSY-11 detection system~\cite{c-11} shown schematically in Fig.~\ref{detector}. 
The target, being a beam of H$_2$ molecules grouped inside clusters of up to 10$^5$ atoms~\cite{dombrowski},
crosses perpendicularly the beam of protons circulating in the ring.
If at the intersection point of the cluster target and COSY beam a collision of protons leads
to the production of a $K^{+}K^{-}$ meson pair, then the reaction products having smaller
momenta than the circulating beam are directed by the magnetic dipole field towards the COSY-11
detection system and leave the vacuum chamber through a thin exit foils~\cite{c-11}. 
Tracks
of positively charged particles, registered by  drift chambers, are traced back
through the magnetic field to the nominal interaction point leading to a momentum 
determination. Knowledge of the momentum combined with 
a simultaneous measurement of the velocity, performed by means of
scintillation detectors S1 and S3, permits to identify the registered particle and to determine
its four momentum vector. Since at threshold the center-of-mass momenta of the
produced particles are small compared to the beam momentum, in the laboratory frame
all ejectiles are moving with almost the same velocity. This means that the laboratory
proton momenta are almost two times larger then the momenta of kaons. Therefore,
in the dipole field protons experience 
a much larger Lorentz force than kaons.
As a consequence, in case of the near threshold production, protons and kaons are
registered in separate parts of the drift chambers. 
Therefore, as a first step in the reaction identification
events with two protons registered in an appropriate 
part of the drift chambers are selected based on the time-of-flight
between the S1 and S3 scintillation hodoscopes. 
The additional requirement that
the mass of the third particle, 
registered at the far side of the chamber with
respect to the circulating beam, 
corresponds
to the mass of the kaon, allows to identify events with 
a $pp\rightarrow ppK^{+}X^{-}$ reaction~\cite{moskal1}. 
\begin{figure}[h]
  \includegraphics[angle=0,width=0.35\textwidth]{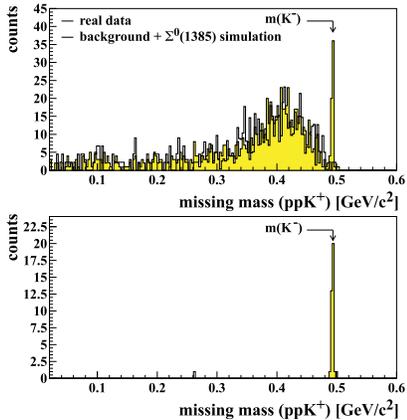}
  \caption{(Upper panel) Exemplary missing mass spectrum determined for the $pp\rightarrow ppK^{+}X^{-}$ reaction 
    at an excess energy of Q = 17 MeV \cite{cosy1}; (Lower panel) Missing mass distribution from upper panel after additional requirement 
    of the signal in the dipole detector as it is described in the text.
    \label{missing}
  }
\end{figure}
Knowing both the four momenta of positively charged ejectiles and the
proton beam momentum one can calculate the mass of an unobserved system $X^{-}$.
Figure~\ref{missing} (upper panel) presents an example of the missing mass spectrum with respect to
the identified $ppK^{+}$ subsystem.
In the case of the $pp\rightarrow ppK^{+}K^{-}$ reaction this 
should correspond to the mass of the $K^{-}$ meson, and indeed a pronounced signal can be clearly seen 
at this position. 
The additional broad structure seen in the upper panel of Fig.~\ref{missing} is partly due to the 
$pp\rightarrow pp\pi^{+}X^{-}$ reaction, where the $\pi^{+}$ was misidentified 
as a $K^{+}$ meson, and in part due to the $K^{+}$ meson production 
associated with the hyperons $\Lambda$(1405) or $\Sigma$(1385)~\cite{cosy1,cosy2}.
The background, however, can be completely reduced by demanding a signal
in the silicon pad detectors (mounted inside the dipole) at the position where the $K^{-}$ meson originating from the
$pp\rightarrow ppK^{+}K^{-}$ reaction is expected (lower panel of Figure 2).
This clear identification allows to select events originating from the $pp\rightarrow ppK^{+}K^{-}$ reaction 
and to determine the total and differential cross sections.\\
Figure~\ref{totalcross} shows the  excitation function 
for the  $pp\rightarrow ppK^{+}K^{-}$ 
reaction established by the  COSY-11~\cite{magnus,cosy1,cosy2}
group near the threshold and 
by ANKE~\cite{anke} and DISTO~\cite{disto} collaborations
at  higher energies.
\begin{figure}
 \includegraphics[height=.3\textheight]{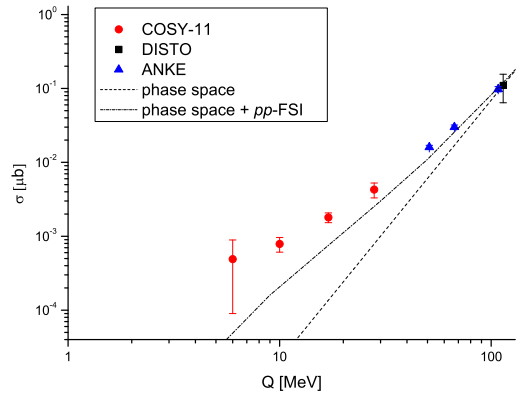}
\vspace{-1cm}
 \caption{Total cross section as a function of the exess energy Q for the 
   $pp\rightarrow ppK^{+}K^{-}$ reaction. 
   The data are form references~\cite{magnus,cosy1,cosy2,anke,disto} 
   and the meaning of the lines is described in the text.
   \label{totalcross}
  }
\end{figure}
The dashed line  shows 
the result of calculations under the assumption of a homogeneous phase 
space population, normalized to the DISTO data point (Q~=~114~MeV).
For the dashed--dotted 
line, the proton-proton final state interaction is included using parameterization
known from the three body final state~\cite{swave}.
It is clearly seen that  calculations neglecting the
interaction of kaons underestimate the
experimental results by a factor of five in the vicinity of the kinematical threshold.
Therefore, the enhancement may be due to the influence of $K^{-}p$ or $K^{+}K^{-}$ 
interaction.
And indeed, with a factorization ansatz for the 
$pp$ and $pK^{-}$ interaction, 
the ANKE 
collaboration described the excitation function much better~\cite{anke,c_wilkin}, 
however still underestimating the two lowest data points by more than a factor of two.
This could indicate that in this region 
the influence of the $K^{+}K^{-}$ interaction is significant and cannot be neglected.
This observation encouraged us to carry out 
the analysis of differencial cross sections
for the low energy data at Q~=~10~MeV (27~events) and Q~=~28~MeV (30 events), 
in spite of the fact that the available statistics is quite low~\cite{cosy2}. 

\section{Goldhaber plot analysis: generalization of the Dalitz plot for four
particle final state}
Usage of the Dalitz plot for extracting information
about the interaction among particles  
in the case of  three body  final states
is well known.
It was introduced by Dalitz in a nonrelativistic application~\cite{dalitz} and then
extended to the relativistic case by Fabri~\cite{fabri}. 
If the transition amplitude is constant over phase space
and if additionally there is no final state interaction, 
the occupation of the Dalitz plot would be fully homogeneous because 
the creation in any phase space interval would be equally probable. 
Thus, final state interaction should show up as a modification 
of the event density in the  Dalitz plot.

\begin{figure}[h!]
 \centering
 \includegraphics[height=.24\textheight]{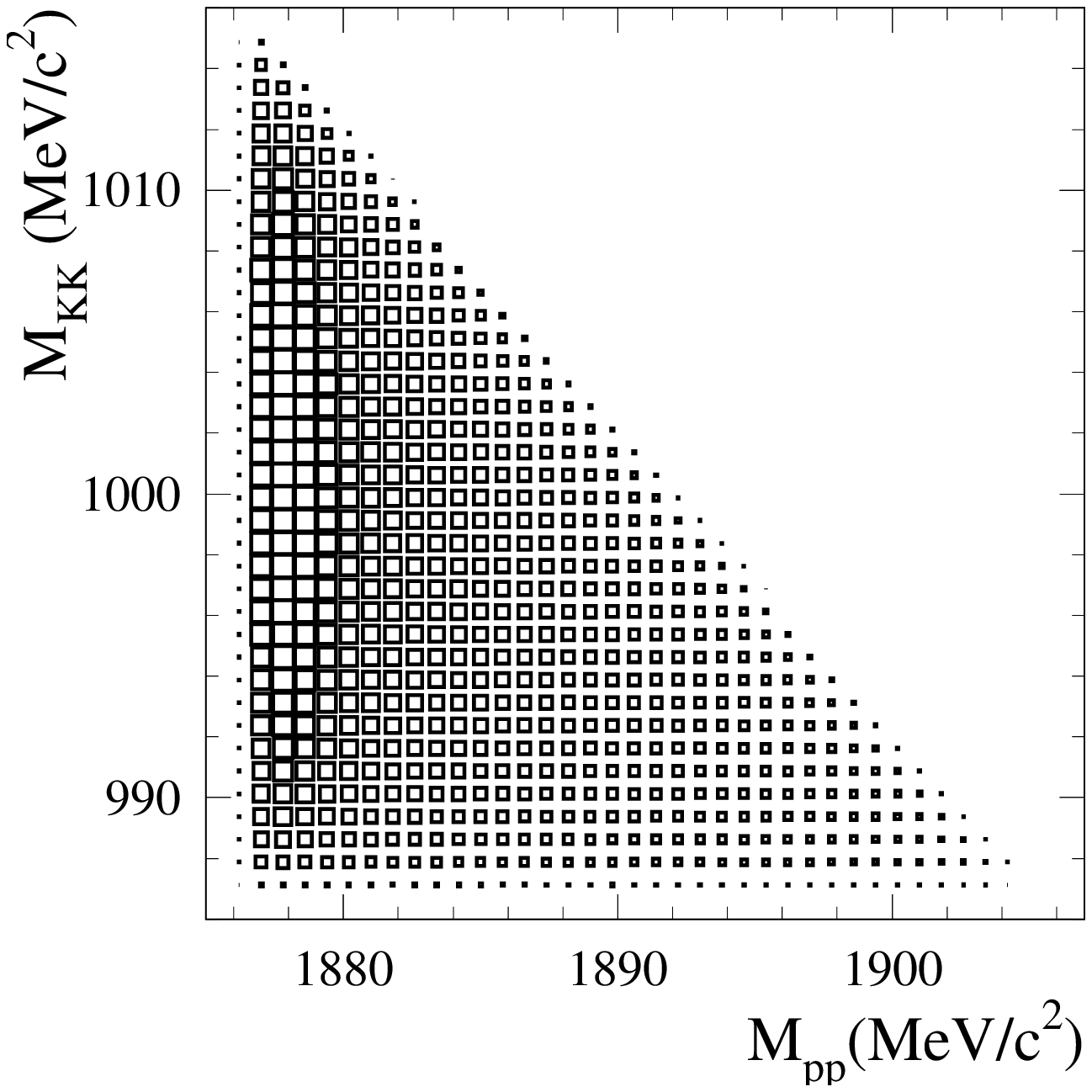}
 \includegraphics[height=.24\textheight]{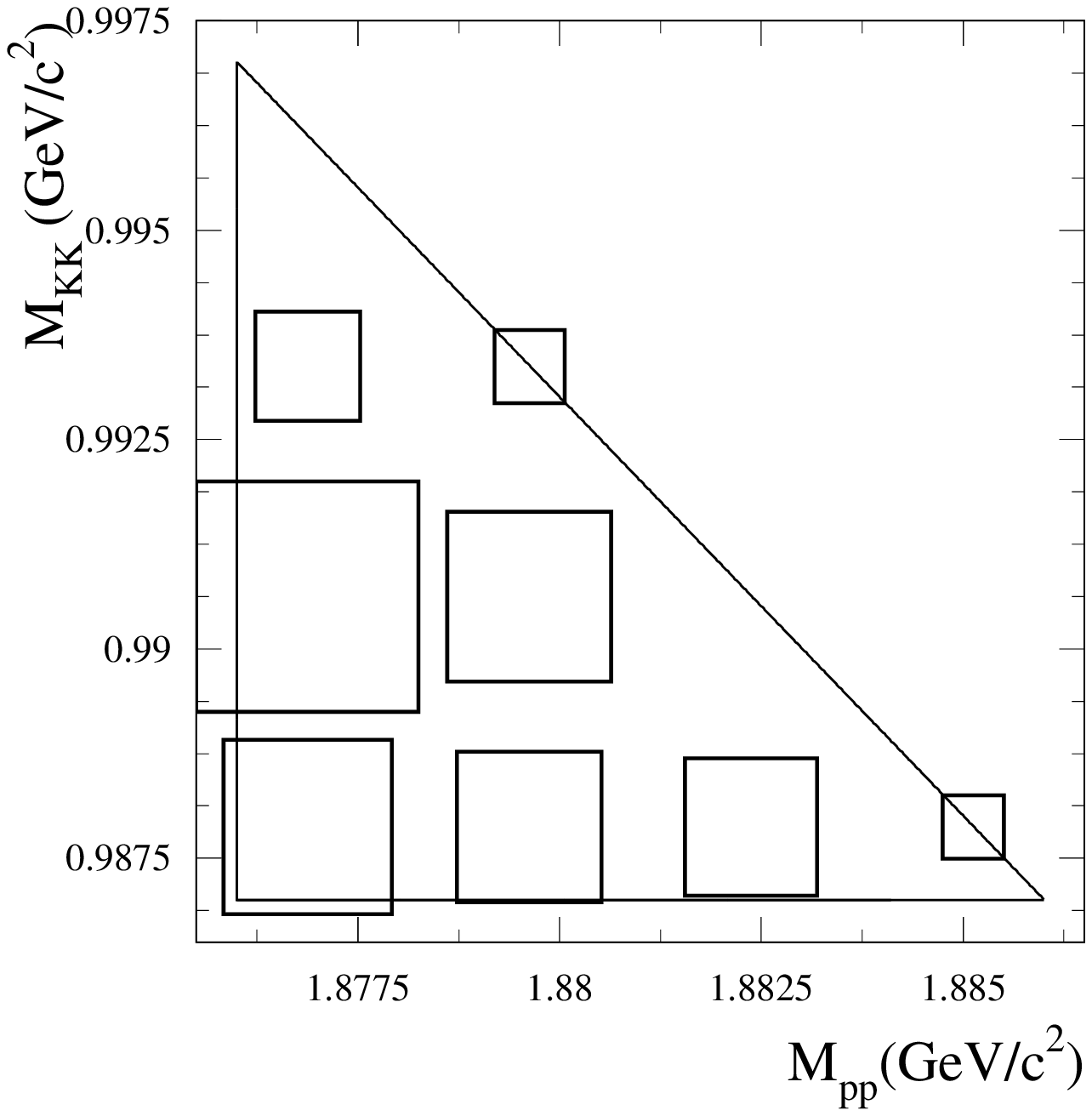}
 \includegraphics[height=.24\textheight]{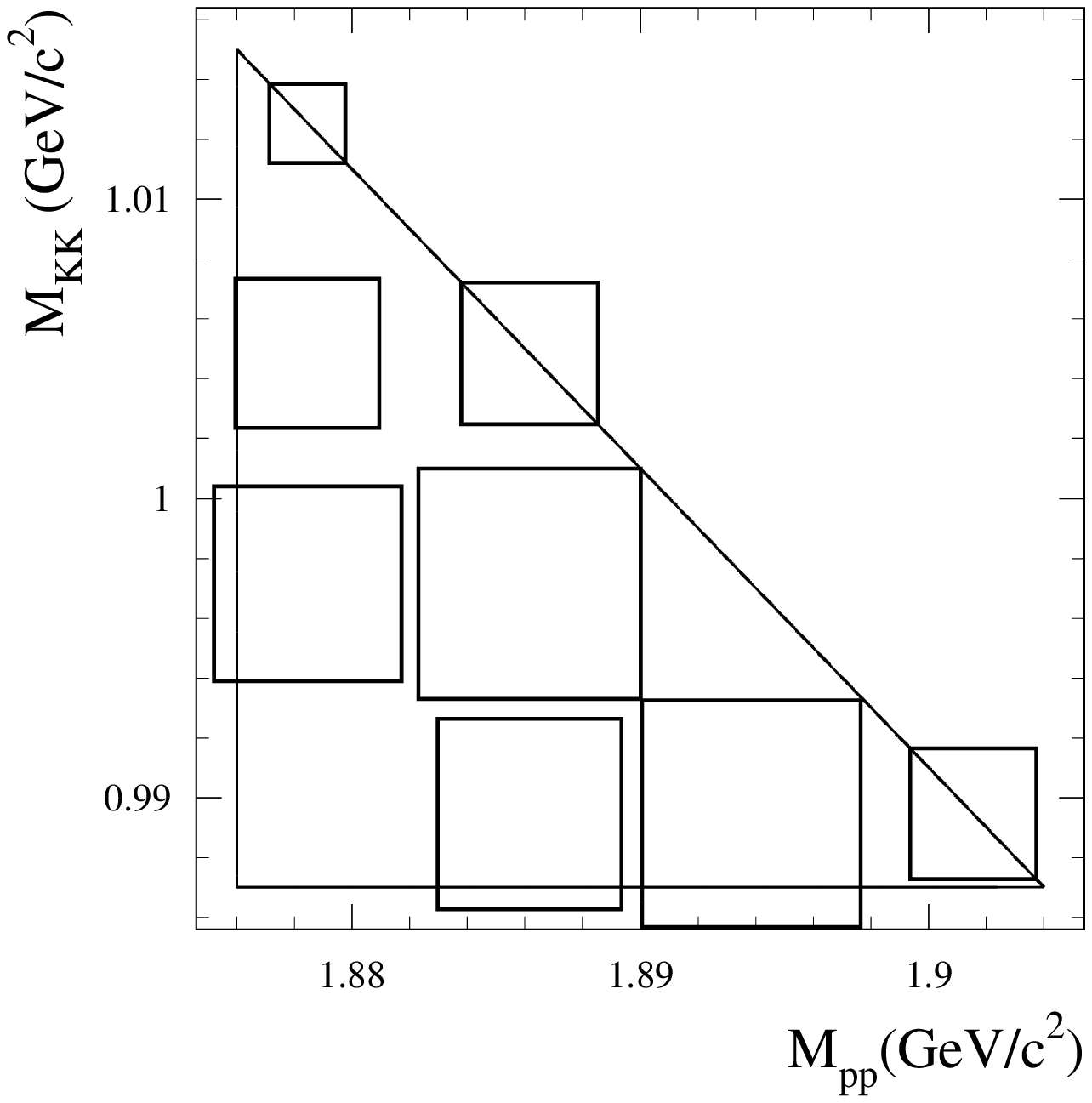}
 \vspace{-0.5cm}
 \caption{Goldhaber plots for the  $pp\rightarrow ppK^{+}K^{-}$ reaction. 
 (upper panel)  Simulations at $Q$~=~28~MeV for 
 homogeneously populated phase space modified by proton-proton final
 state interaction 
 taken into account as weights proportional
 to the inverse of a squared Jost-function of the Bonn potential~\cite{swave,druzhinin}.
 (middle and lower panels)  Experimental data obtained at Q~=~10~MeV (middle), and Q~=~28~MeV (lower).
   \label{experiment}
 }
\end{figure}

In the case of four particles in final state the analysis is more complex, 
because one needs five variables to 
fully describe a relative  movement of particles.
Nevertheless, there are many different types of generalization 
of the Dalitz plot 
for four-body final states. 
In this contribution we present a
generalization proposed by Goldhaber~\cite{goldhaber1,goldhaber2},
which we use further on for studying the interaction in the $ppK^{+}K^{-}$ system.
However, there exist many other approaches as described e.g.\ 
by Nyborg or Chodrow~\cite{nyborg,chodrow,silar}.
Consider a reaction yielding in 
the final state four particles with masses $m_{i}$ and total energy $\sqrt{s}$ 
in the centre-of-mass frame. 
Assuming that the matrix element for the process $M$ depends only on invariant masses
 of two- and three particle subsystems~\cite{nyborg} 
 the distribution of events can be 
 expressed 
 in some choice of five independent invariant masses
e.g.: 
$M_{12}^{2},~M_{34}^{2},~M_{14}^{2},~M_{124}^{2},~M_{134}^{2}$. 
Assuming further that $M$ depends at most on $M_{12}^{2},~M_{34}^{2}$, and $M_{124}^{2}$, 
which corresponds to the situation where only  two two-particle or one three-particle resonances 
are present~\cite{nyborg},
one obtains the following  distribution of events:
\begin{equation}
  d^{3}P=\frac{\pi^{3}\left|M\right|^{2}g\left(M_{12}^{2},m_{1}^{2},m_{2}^{2}\right)}{8sM_{12}^{2}}dM^{2}_{12}dM^{2}_{34}dM^{2}_{124}
\end{equation}
where g is a simple analytical function of its variables~\cite{nyborg}.
The projection of the physical region 
on the $\left(M_{12},~M_{34}\right)$-plane 
gives a right isosceles triangle in which the area is not proportional 
to the phase space volume. 
It is important to note that the event density in the Goldhaber plot 
is not homogeneous and goes to zero on the entire boundary 
of the plot given by the following equations: $M_{12}+M_{34}=\sqrt{s},~M_{12}=m_{1}+m_{2},~M_{34}=m_{3}+m_{4}~$\cite{nyborg}.

Figure~\ref{experiment}~(upper panel) presents  the simulated distribution
for the $ppK^{+}K^{-}$ reaction 
determined taking into account
only 
the pp--FSI. 
Experimental 
event distributions after acceptance corrections 
for both studied excess energies are shown in the middle and lower panels. 
Clearly  the event 
densities in the experimental Goldhaber plots differ from 
the simulated spectrum.
In particular data show an enhancement in the range of  
small $K^{+}K^{-}$ invariant masses which may signify  
a signal from the kaon-antikaon interaction.
A similar enhancement 
is seen  by the ANKE group~\cite{anke} below the $\phi$ meson mass in the  $K^{+}K^{-}$ invariant mass distributions.
As a next 
step in the analysis we will compare 
experimental data to the results of Monte Carlo simulations generated 
with various parameters of the $K^{+}K^{-}$ interaction taking into account the $pK$--FSI, 
as  described in references~\cite{anke,c_wilkin}. 

\section{Acknowledgements}
We acknowledge the support by the
European Community-Research Infrastructure Activity
under the FP6 programme (Hadron Physics,
RII3-CT-2004-506078), by
the Polish Ministry of Science and Higher Education under the grants
No. 3240/H03/2006/31  and 1202/DFG/2007/03,
and by the German Research Foundation (DFG).

\end{document}